\documentclass[useAMS,usenatbib]{mn2e}
\usepackage{amssymb,amsmath,epsfig,times, natbib,color,threeparttable,booktabs}
\pdfoutput=1

\title[NGC 1600 mass concentration]{The unusually high dark matter concentration of the galaxy group NGC 1600}
\author[J. Runge] {\parbox[]{6.5in}{
      J. Runge$^1$\thanks{Email: 
    jmr0062@uah.edu}, {S. A. Walker$^1$}, {M. S. Mirakhor$^1$}}\\
     \footnotesize
     $^1$Department of Physics and Astronomy, The University of Alabama in Huntsville, 301 Sparkman Drive NW, Huntsville, AL 35899, USA \\
}

\date{}

\begin{document}

\maketitle

\begin{abstract}
We investigate the properties of the dark matter (DM) halo surrounding the nearby galaxy group NGC 1600. Through the use of deep (252 ks) \textit{Chandra} observations and 64.3 ks of \textit{XMM--Newton} observations, we construct surface brightness profiles in multiple energy bands in order to perform hydrostatic equilibrium analysis of the hot plasma within NGC 1600. Regardless of the DM model profile assumed, we measure a halo concentration (c$_{200}$) that is an extreme, positive outlier of the $\Lambda$CDM c$_{200}$-M$_{200}$ relation. For a typical NFW DM profile, we measure c$_{200}\!=\!26.7\pm1.4$ and M$_{200}\!=\!(2.0\pm0.2)\times10^{13}$ M$_\odot$; assuming a similar halo mass, the average concentration expected is c$_{200}=6-7$ for the theoretical $\Lambda$CDM c-M relation. Such a high concentration is similar to that of well-known fossil groups MRK 1216 and NGC 6482. While NGC 1600 exhibits some properties of a fossil group, it fails to meet the X-ray luminosity threshold of L$_X>5\times10^{41}$ erg s$^{-1}$. Whether or not it is considered a fossil group, the high concentration value makes it part of a select group of galaxy groups.
\end{abstract}

\section{Introduction}
In the hierarchical formation of structures in the universe, smaller structures form first before merging to become larger structures. Following this logic, galaxy groups form before clusters. If a group is sufficiently isolated and formed early enough, merging of the most massive galaxies in the group can result in a system that is dominated by a large, central galaxy. Surrounding this central galaxy are faint, less massive galaxies and an X-ray halo. Such systems are referred to as `fossil' groups (\citealt{Ponman1994}).

Since their introduction by \cite{Ponman1994}, studies have revealed how the properties of fossil groups differ from those of other galaxy groups. A study by \cite{Jones2003} showed that fossil groups have much higher X-ray luminosities than those of normal groups; possibly attributed to low central gas temperature or low entropy. This discovery along with the previously mentioned massive, central galaxy determine what we now consider to be the main characteristics of fossil groups: an X-ray luminosity of L$_X>5\times10^{41}$ erg s$^{-1}$ and an R-band magnitude gap of at least 2 between the brightest, central galaxy and the next brightest galaxy (\citealt{Harrison2012}).

The bright x-ray halo surrounding the central galaxy is indicative of a group-sized dark matter (DM) halo. Through various X-ray surveys of galaxy groups, clusters, and large elliptical galaxies (e.g.\citealt{Buote2007,Schmidt2007,Ettori2010}), there is a well established power-law relationship between the concentration (c) and mass (M) of a system that is consistent with $\Lambda$CDM cosmology. It has been asserted that the concentration is set by the mean density of the universe when the halo was formed (\citealt{Navarro96}); with halos that formed earlier having higher concentrations. Due to their early formation time, fossil groups should appear as positive outliers in the c-M relation, with higher than expected concentration values. X-ray observations of a number of fossil groups has shown that they do indeed have unusually high concentrations of c$_{200}\approx10-30$. (\citealt{Humphrey2011,Humphrey2012,Buote2017,Buote2019}).

In their paper, \cite{Smith2008} discussed the possibility of NGC 1600 being part of a fossil group. It is much brighter than its neighbors within the group and there is an X-ray halo, but it falls below the X-ray luminosity limit of other fossil groups. In this paper, we make use of 252 ks of \textit{Chandra} X-ray observations and 64.3 ks of \textit{XMM--Newton} observations in order to construct surface brightness profiles in multiple bands stretching out 77 kpc from the center of NGC 1600. By fitting these radial profiles using \texttt{MBPROJ2} to known mass profiles, we hope to put constraints on the mass profile and concentration. These measurements should reveal whether it is an outlier on the c-M relation and deserves to be labeled a fossil group. 

Throughout the paper, we assume a $\Lambda$CDM cosmology with $\Omega_\Lambda=0.7$, $\Omega_m=0.3$, and $H_0=70$~km s$^{-1}$ Mpc$^{-1}$. At the redshift of NGC 1600, 1 arcsecond corresponds to 0.32 kpc. All errors are 1-$\sigma$ unless otherwise noted.

\begin{figure*}
	\includegraphics[width=\textwidth]{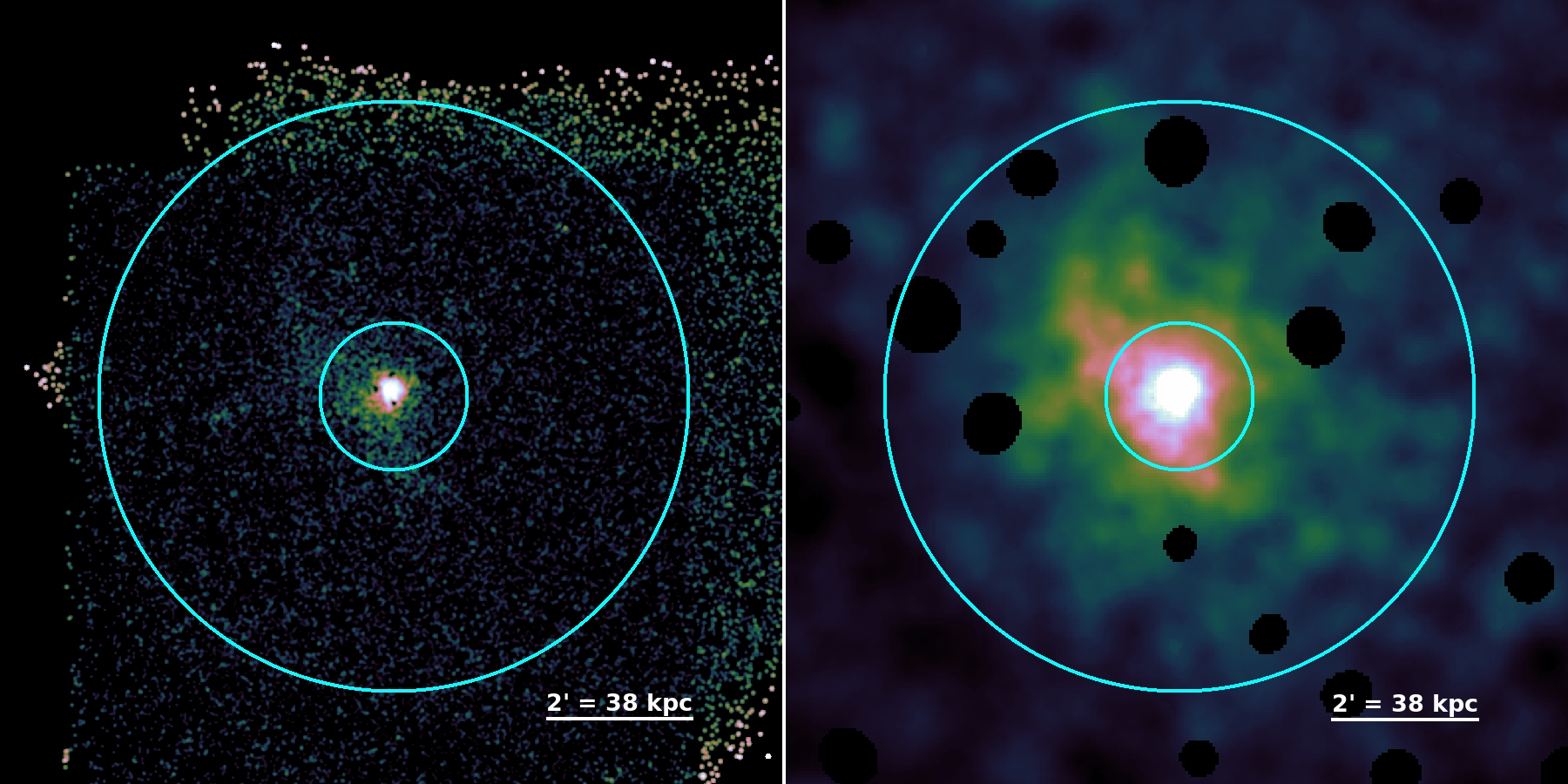}
	\caption{Left: \textit{Chandra} soft-band (0.5-1.2 keV), exposure-corrected flux image for NGC 1600. Right: Same as left for XMM-Newton. Point sources have been removed in each image. The inner region of radius $1'$ denotes the area where \textit{Chandra} data was used to create surface brightness profiles. From the inner radius to the outer radius of $4'$ is the region where XMM-Newton data was used to create surface brightness profiles.}
	\label{fig:image_Chandra_XMM}
\end{figure*}

\section{Data}
\subsection{\textit{Chandra}}
We make use of the same deep \textit{Chandra} X-ray observations used in our previous paper \cite{Runge2021} for fine detail within the core. The eight observations used are outlined in Table~\ref{table:observations}, six of our own taken between 2018 and 2019 and two archival observations taken in 2002. In total, we have 252 ks of observation time. All data were reprocessed using the \texttt{chandra\_repro} script of \texttt{CIAO 4.13} and \texttt{CALDB 4.9.1}. Point sources were detected using \texttt{CIAO wavdetect} on the merged broad energy image and excluded from analysis (Left Fig.~\ref{fig:image_Chandra_XMM}).

\subsection{\textit{XMM--Newton}}
To extend our data beyond the core, we make use of \textit{XMM--Newton}'s larger FOV and effective area. NGC 1600 was observed with the European Photon Imaging Camera (EPIC) abroad \textit{XMM--Newton} for around 64.3 ks between 2007 February 6 and 2007 February 7 (Table \ref{table:observations}). The data were reduced using the \textit{XMM--Newton} Science Analysis System (\textit{XMM}-SAS) version 18.0 and current calibration files (CCF), following the procedure illustrated in the Extended Source Analysis Software (ESAS) cookbook\footnote{https://heasarc.gsfc.nasa.gov/docs/xmm/esas/cookbook/xmm-esas.html}. 

We initiated the data processing by running the \textit{epchain} and \textit{emchain} tasks, followed by the \textit{mos-filter} and \textit{pn-filter} tasks to filter the data for soft proton flares and create clean event files for the MOS and PN detectors. We then examined the data of the MOS detectors for CCDs in anomalous states, and any affected CCDs were excluded from further analysis. Point sources and extended substructures that contaminated the field of view were detected and removed by running the ESAS source-detection tool \textit{cheese}.  For all detectors, we also run \textit{mos-spectra} and \textit{pn-spectra} to create spectra, response matrix files (RMFs), ancillary response files (ARFs), and exposure maps for the interested region. These files were then used to create the quiescent particle background spectra and images in the MOS and PN coordinates by running the \textit{mos-back} and \textit{pn-back} scripts. Furthermore, we filtered the data for any soft proton contamination that may have remained after the initial light-curve screening by running the \textit{proton} task. 

The analysis procedure described above created all of the primary components required for a background-subtracted and exposure-corrected image. After weighting the \textit{XMM--Newton} detectors by their effective area, these primary components from the MOS and PN detectors were merged and adaptively smoothed into a single image. The data were also examined for any remaining point sources that were missed using \textit{cheese} by running the \textit{Chandra} source-detection tool \textit{wavdetect} with wavelet scales of 2, 4, 8, 16, and 32 pixels, and any detected point sources were then excluded from further analysis (Right Fig.~\ref{fig:image_Chandra_XMM}).

\section{Analysis}

\subsection{Surface Brightness Profiles}
Surface brightness profiles were extracted using the optical peak as the center. In order to obtain profiles for various energy ranges, we created images in eight energy bands between adjacent energies of 0.5, 0.75, 1, 1.25, 1.5, 2, 3, 4, and 5 keV. Radial profiles of counts, area, exposure, and background rates were extracted for each energy range. \textit{Chandra} data was used for the inner $1'$ and \textit{XMM--Newton} data was used from $1'$ out to $4'$ (Shown in Fig.~\ref{fig:image_Chandra_XMM}). The XMM data was appropriately scaled to \textit{Chandra} in order to account for the difference in effective area. The surface brightness profile for energy range 1.0 to 1.25 keV is shown in Fig~\ref{fig:Sb_profile}.

\begin{table}
    \centering
     \caption{Overview of the \textit{Chandra} and \textit{XMM-Newton} observations used in this study.}
    \begin{threeparttable}
        \begin{tabular}{c r|c|c}
     \hline
     \hline
     Observatory & Obs. ID & Date & Exposure (ks)  \\
      \hline
      \vspace{-1 em} \\
       \textit{Chandra}&4283  & 18-09-02  & 28.5  \\
        &4371  & 20-09-02  & 28.5   \\
        &21374 & 03-12-18  & 26     \\
        &21375 & 28-11-19 & 42     \\
        &21998 & 03-12-18  & 14    \\
        &22878 & 25-11-19 & 46     \\
        &22911 & 01-11-19  & 31    \\
        &22912 & 02-11-19  & 36     \\
        \vspace{-1 em} \\
        \textit{XMM-Newton}&0400490201 & 07-02-07 & 64.3 \\
      \hline
      \end{tabular}
     \label{tab:obs}
     \end{threeparttable}
        
    \label{table:observations}
\end{table}

\subsection{Profile Modeling}
We use the program \texttt{MBPROJ2}, discussed in detail in \cite{Sanders2018}, since it allows for fine detail with a small number of counts. An added benefit is that by utilizing the multiband, projected surface brightness profiles and assuming hydrostatic equilibrium, \texttt{MBPROJ2} is able to compute the deprojected pressure, density, and temperature profiles. The concentration parameter, c$_{200}$, is found from the fitting routine based upon the dark matter halo profile.

\begin{figure}
	\includegraphics[width=0.49\textwidth]{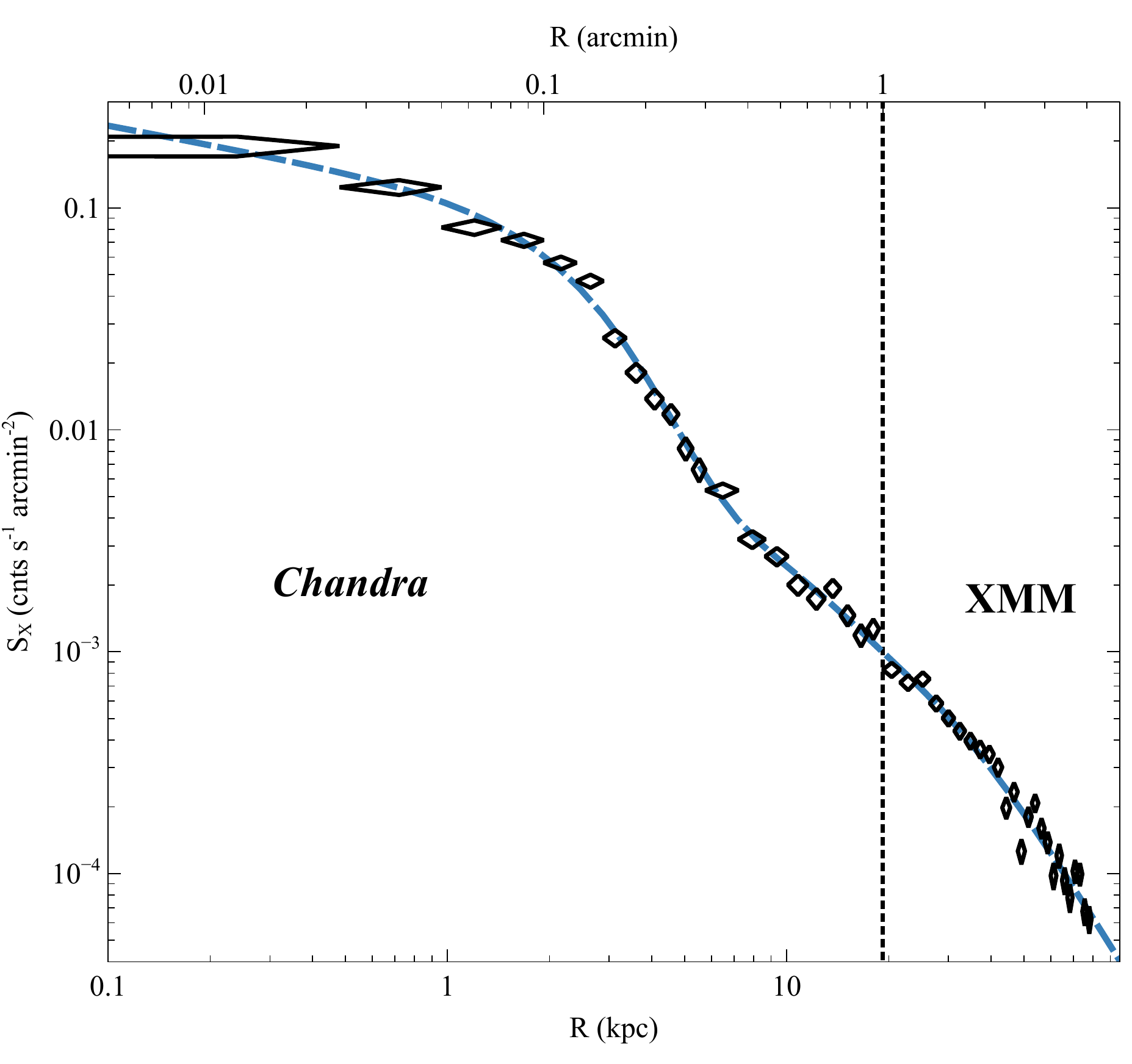}
		\caption{The surface brightness profile for the 1.0 to 1.25 keV energy range. The dotted line at $1'$ (19.2 kpc) marks the boundary between the \textit{Chandra} and XMM data. The dashed blue line shows the modified-$\beta$ profile fit.}
	\label{fig:Sb_profile}
\end{figure}

To test the effects of different models, we use two different models to describe the dark matter halo: the Navarro-Frenk-White (NFW) profile (\citealt{NFW97}) and the Einasto profile (\citealt{Einasto65}). The mass density profile for the NFW profile (\citealt{NFW97,Schmidt2007}) is described by:
\begin{equation}
\rho(r) = \frac{\rho_0}{r/r_{\textup{s}}(1+(r/r_{\textup{s}}))^2}
\end{equation}
\begin{equation}
\rho_0 = 200\rho_cc^3_{200}/3(ln(1+c_{200})-c_{200}/(1+c_{200}))
\end{equation}
where $\rho_c$ is the critical density of the universe, and $\textup{c}_{200} =\textup{r}_{200} /\textup{r}_s$. In contrast, the Einasto profile is given by:
\begin{equation}
\rho(r)=\rho_s ~\textup{exp}\left ( -\frac{2}{\alpha} \left [ \left ( \frac{r}{r_s} \right )^\alpha - 1 \right ]\right )
\end{equation}
In this study we fix $\alpha=0.16$, which is applicable for NGC 1600's halo mass (e.g. figure 13 of \citealt{Dutton2014}). 
We use a modified-$\beta$ profile (\citealt{Vik2006}) to describe the electron density:
\begin{equation}
n_{e}^2=n_{0}^2 \frac{\left ( r/r_c \right )^{-\alpha}}{\left ( 1+r^2/r_c^2 \right )^{3\beta-\alpha/2}}\frac{1}{\left ( 1+r^\gamma/r_s^{\gamma} \right )^{\varepsilon/\gamma}}+\frac{n_{02}^2}{\left ( 1+r^2/r_{c2}^2 \right )^{3\beta_2}}
\end{equation}
where all parameters follow the constraints outlined in $\S$2.5.5 of \cite{Sanders2018}. As Fig~\ref{fig:Sb_profile} shows, this provides a good fit to the data.

While the NFW profile is included in \texttt{MBPROJ2}, the Einasto profile is not; therefore, we modified the code to account for this deficit. \texttt{MBPROJ2} makes use of a Markov chain Monte Carlo (MCMC) routine for fitting. The parameters we choose for this routine are: a chain length of 1000, a burn-in period of 1000 steps, and 200 walkers. The Galactic column density was fixed at a value of $N_{\textup{H}}=3.14\times10^{20}$ cm$^{-2}$ (\citealt{DL90}).
\subsubsection{Stellar and Black Hole Mass}
When determining the DM halo, we want to account for both the stellar mass ($8.3\times10^{11}$M$_\odot$; \citealt{Ma2014,Thomas2016}) and the black hole mass ($17\pm1.5\times10^9$ $\textup{M}_{\odot}$; \citealt{Thomas2016}). A point source with a mass matching that of the black hole is used to see the effect the black hole has on the DM halo. The stellar mass is incorporated using two different methods. The first method uses a Sersic profile based upon the effective radius found by \citealt{Ma2014} using the K-band magnitude from the Extended Source Catalog (XSC; \citealt{Jarrett2000}) of the Two Micron All Sky Survey (2MASS; \citealt{Skrutskie2006}). The second method directly scales the surface brightness profile of the K-band 2MASS image to a density profile.

\section{Results}

\subsection{Concentration}
\begin{table*}
    \centering
        \begin{tabular}{c r c c c c c c c  c c}
        \toprule
        &&\multicolumn{4}{c}{NFW} && \multicolumn{4}{c}{Einasto}\\
        \cmidrule{3-6} \cmidrule{8-11}
        &&\multicolumn{2}{c}{$c_{200}$} & \multicolumn{2}{c}{$r_{200}$ (kpc)}&&\multicolumn{2}{c|}{$c_{200}$} & \multicolumn{2}{c}{$r_{200}$ (kpc)}\\
        Data&& Best & $\Bar{c}_{200}$ & Best & $\Bar{r}_{200}$&& Best & $\Bar{c}_{200}$ & Best & $\Bar{r}_{200}$\\
        \toprule
    \textit{Chandra}&Single     & 70.8 & $69.7\pm6.4$ & 333.4 & $337\pm17$&& 42.7 & $43.1\pm4.2$ & 427.6 & $432\pm29$ \\
    &Fixed BH     & 27.5 & $27.7\pm1.7$ & 590.2 & $589\pm29$&&21.9&$22.2\pm2.5$&663.7&$656\pm60$ \\
    &Sersic Profile & 60.3 & $60.3\pm5.5$ & 353.2 & $353\pm20$&& 45.7 & $45.5\pm4.9$ & 391.7& $397\pm29$ \\
    &2MASS Profile  & 72.4 & $72.9\pm6.7$ & 325.8 & $325\pm16$&&46.8&$45.4\pm5.4$&397.2&$410\pm33$ \\
    &BH + Sersic     & 26.9 & $27.1\pm1.9$ & 558.5 & $560\pm31$&&18.6&$18.7\pm1.7$&679.2&$675\pm47$ \\
    \vspace{-1em}\\
    \textit{Chandra}+XMM &Single & 29.5 & $29.8\pm1.6$ & 626.6 & $623\pm27$&&20.9&$21.5\pm1.5$&729.5&$721\pm39$\\
    &Fixed BH     & 24.0 & $23.8\pm1.3$ & 698.2 & $699\pm32$&&17.4&$17.6\pm1.0$&790.7&$776\pm33$ \\
    &Sersic Profile & 28.2 & $28.3\pm1.7$ & 615.2 & $610\pm30$&& 22.9 & $23.8\pm1.5$ & 642.7 & $630\pm29$ \\
    &2MASS Profile  & 35.5 & $37.0\pm2.3$ & 521.2 & $511\pm23$&&29.5&$29.3\pm2.1$&545.8&$555\pm29$ \\
    &BH + Sersic     & 23.4 & $23.9\pm1.2$ & 647.1 & $636\pm24$&&16.6&$17.0\pm1.1$&758.6&$746\pm36$ \\
    &Decreasing Z  & 31.6 & $31.9\pm1.8$ & 567.5 & $570\pm25$&&24.5&$24.6\pm1.8$&633.9&$631\pm34$ \\
    &Decreasing Z + BH + Sersic & 26.3 & $26.7\pm1.4$ & 564.9 & $557\pm21$&&20.0&$19.7\pm1.3$&630.1&$638\pm29$ \\
    \toprule
        \end{tabular}
        \caption{Fitting results of $c_{200}$ and $r_{200}$ for NGC 1600 using an NFW profile and Einasto profile. We present both the best fitting result along with the median value and 1$\sigma$ scatter. }
    \label{table:fitting}
\end{table*}
The fitting routine in \texttt{MBPROJ2} returns the best-fit values for c$_{200}$ and r$_{200}$ as well as the distribution of these values from the MCMC routine. We use the distribution to evaluate the median value and 1$\sigma$ scatter for each parameter. The results of the fitting are shown in Table~\ref{table:fitting}. With just the \textit{Chandra} data and not accounting for the influence of stellar mass and central black hole, we find extremely large values for c$_{200}$. Once everything is incorporated, the concentration falls to c$_{200}\!=\!27.1\pm1.9$ and c$_{200}\!=\!18.7\pm1.7$ for the NFW and Einasto profiles respectively. We choose to use the Sersic profile for the final fitting in order to get the most conservative value for c$_{200}$ possible.

\begin{figure}
	\includegraphics[width=0.49\textwidth]{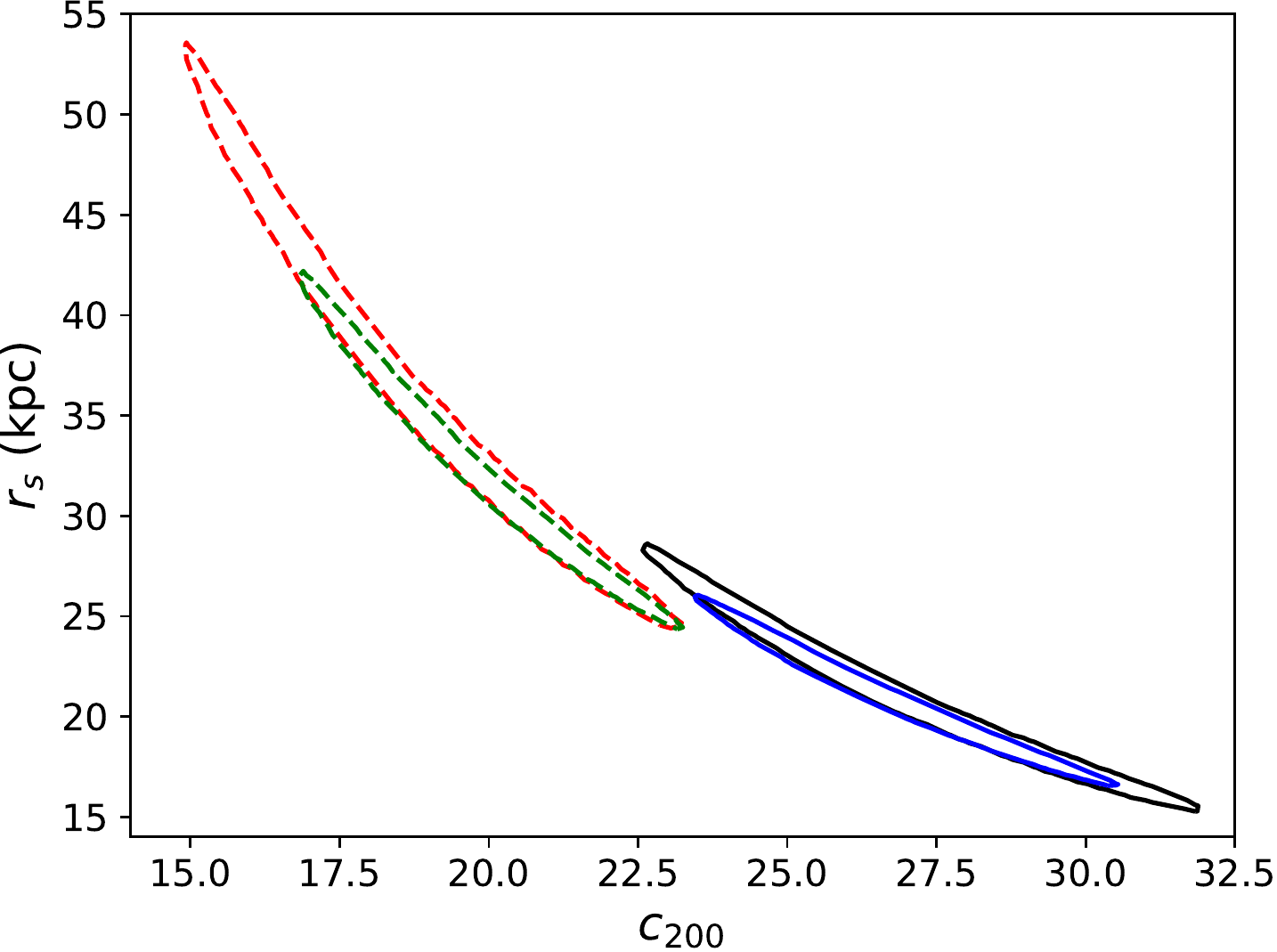}
	\caption{The $2\sigma$ confidence contours for the NFW (solid black) and Einasto (dashed red) fitting. Contours found with the inclusion of the XMM data are also shown (solid blue for NFW and dashed green for Einasto). }
	\label{fig:concentration_NFW_Ein_contours}
\end{figure}

With the addition of the XMM data, we need to adjust our fitting parameters appropriately. For galaxy groups, the metal abundance is relatively constant near the center but decreases with radius further out (\citealt{Humphrey2006,Rasmussen2009,Mernier2017}). Therefore, while our assumption of a constant metal abundance when using just the \textit{Chandra} data may have been justified, we must account for the decreasing metal abundance when including the XMM-Newton data. We accomplished this by setting the metal abundance to Z/Z$_\odot$=1 for the inner $1'$ and steadily decreasing to a value of Z/Z$_\odot$=0.3 out to $4'$. The final concentration values, found by varying the metal abundance and including the black hole and stellar mass, are c$_{200}=26.7\pm1.4$ and c$_{200}=19.7\pm1.3$ for the NFW and Einasto profiles respectively (See Table~\ref{table:fitting}). The addition of the \textit{XMM--Newton} data results in comparable concentration values found using only \textit{Chandra}, $<\! 1.5$ difference, but with slightly improved uncertainties. A comparison of the $2\sigma$ confidence contours when including the XMM data is shown in Fig.~\ref{fig:concentration_NFW_Ein_contours}.

\section{Discussion}
\subsection{High Concentration}
To better understand why the concentration value is such an outlier, we must discuss the concentration-mass relation. The halo concentration and mass (M$_{200}$) have been shown to have a relation that evolves with redshift and is typically characterized as a power law (i.e. $c_{200} \propto M_{200}^{a}$) with a shallow slope, $a\!\approx-1$ (\citealt{Navarro96,Bullock2001,Buote2007,Bhattacharya2013,Dutton2014}). Assuming spherical symmetry, M$_{200}$ is simply found by
\begin{equation}
M_{200} = \frac{4}{3}\pi200\rho_cr_{200}^{3}
\end{equation}
We calculate M$_{200}$ for both the NFW and Einasto profile: $(2.0\pm0.2)\times10^{13}$ and $(3.0\pm0.4)\times10^{13}$ M$_\odot$ respectively. 

Given the halo mass of NGC 1600, \cite{Dutton2014} predict that it should have a much lower concentration of $\sim$6--7. This puts it in similar company with two other known outliers: NGC 6482 and MRK 1216; whose concentrations are $c_{200}=30.4\pm4.3$ and $c_{200}=32.2\pm7.1$ respectively (\citealt{Buote2017, Buote2019}). NGC 1600 has a lower c$_{200}$ than either NGC 6482 and MRK 1216, but it also has a larger mass, compared to M$_{200}\sim\!5\times10^{12}$ M$_\odot$ for NGC 6482 and MRK 1216. This follows the trend that c$_{200}$ decreases as M$_{200}$ increases as predicted by CDM models (\citealt{Navarro96,Macci2007}). 

\subsection{Fossil group}
\begin{figure*}
	\includegraphics[width=0.49\textwidth]{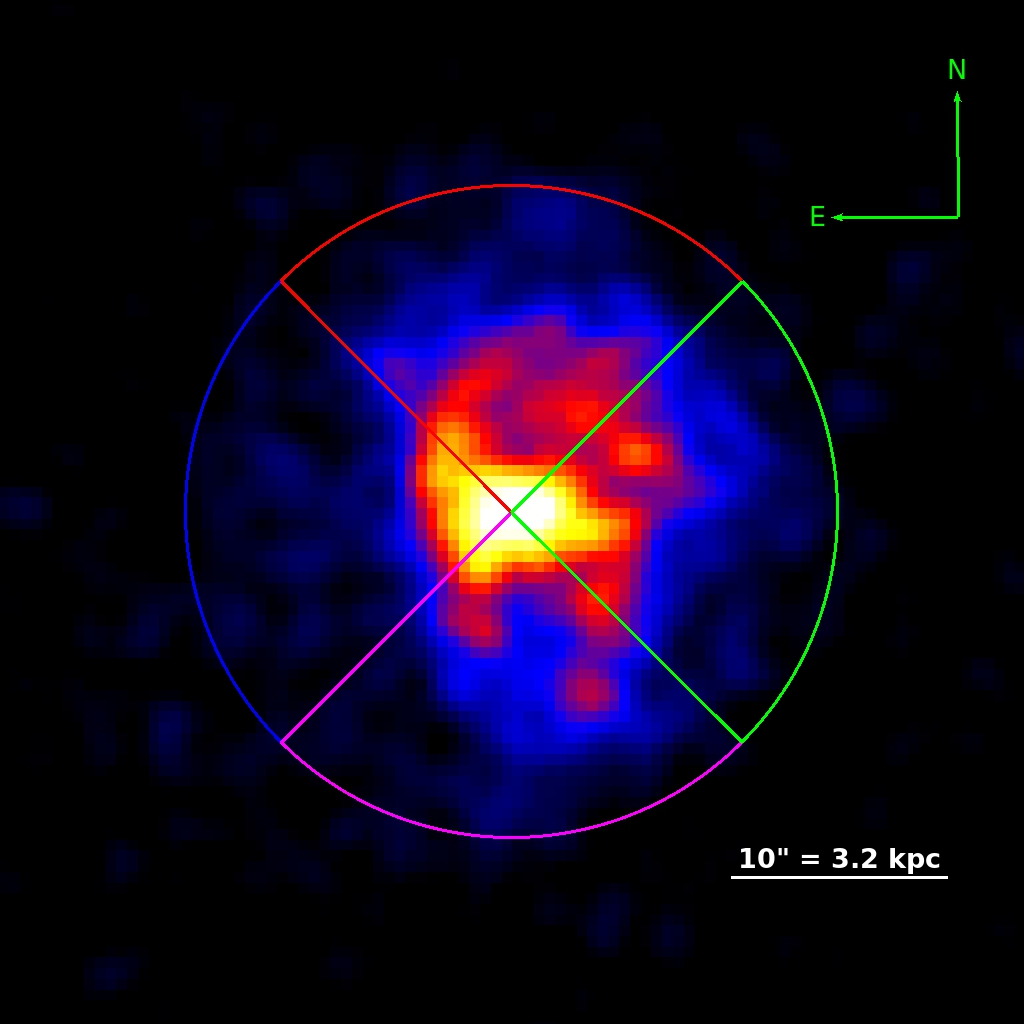}
	\includegraphics[width=0.49\textwidth]{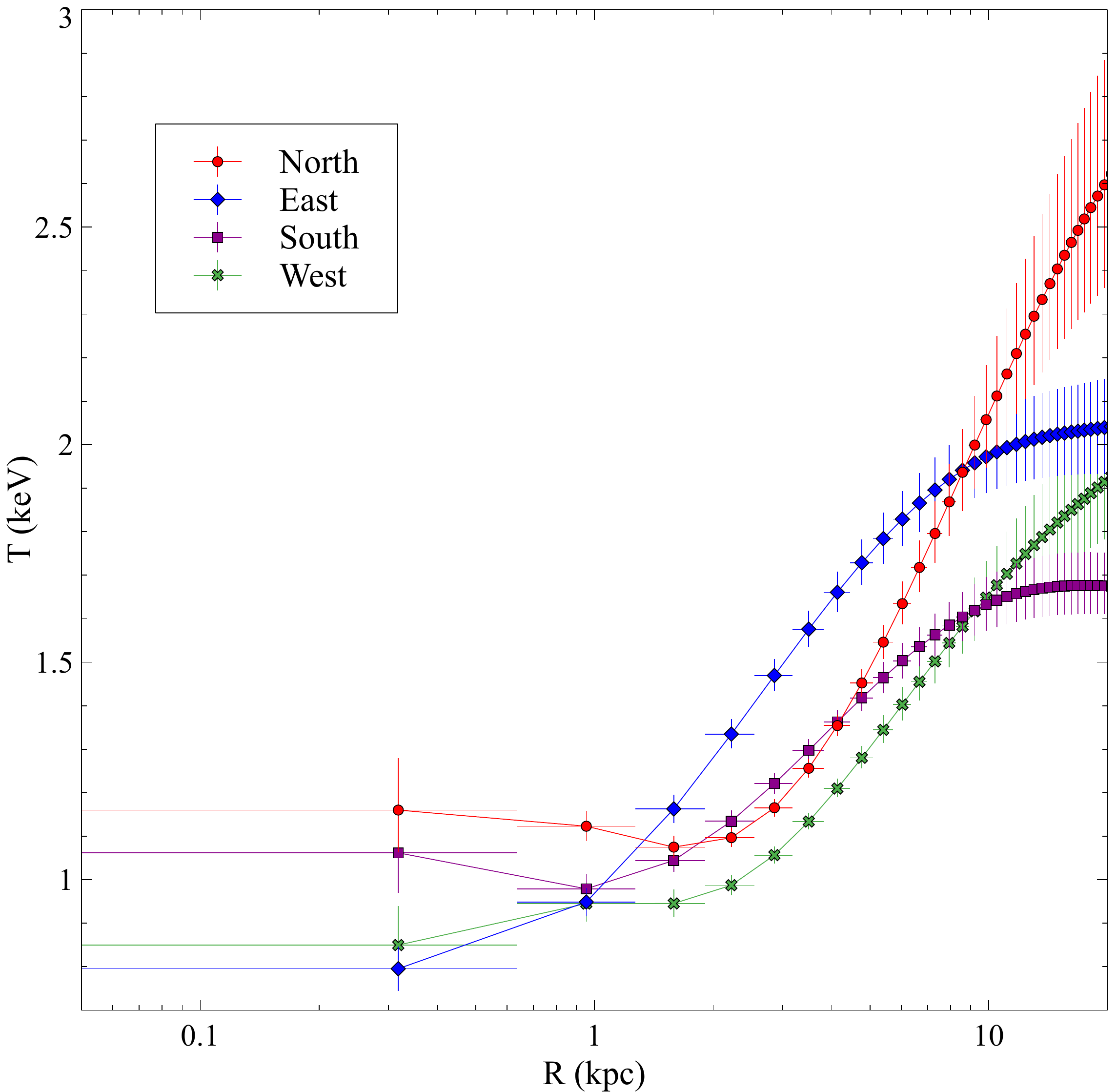}
	\caption{Left: Soft-band image of NGC 1600 with the four quadrants (N, E, S, and W) outlined. Right: Comparing the temperature profiles for the quadrants shown on the left.}
	\label{fig:image_temp_compare}
\end{figure*}
For both NGC 6482 and MRK 1216, the high concentration is associated with the fact that both systems are considered fossil groups. Fossil groups are thought to have early formation times which is reflected in their high concentration values; a property supported by \cite{NFW97} which argued that the concentration is determined by the mean density of the universe at the redshift of formation (i.e. higher concentration means earlier formation).

As discussed in \cite{Smith2008}, NGC 1600 has a lot of similar characteristics to that of known fossil groups: a large magnitude difference between it and its neighbors, extended X-ray emission, and much of its mass was already in place 7--8 Gyrs ago. However, NGC 1600 has an X-ray luminosity of L$_X = 2.9\times10^{41}$ erg s$^{-1}$ (\citealt{Sivakoff2004}), slightly lower than the luminosity typically associated with a fossil group (L$_X > 5\times10^{41}$ erg s$^{-1}$). \cite{Smith2008} suggest that NGC 1600 is a new class of fossil group, one that formed through the merger of a poor group, thus resulting in a lower X-ray emission. It is also possible that powerful feedback from the $17\times10^{9}$ M$_\odot$ black hole has forced gas outwards, lowering the density and X-ray luminosity. For comparison, stellar kinematics show that MRK 1216 contains a SMBH of $M_{\text{BH}}=4.9\pm1.7\times10^9$M$_\odot$ (\citealt{Walsh2017}). NGC 6482 may contain a SMBH with $M_{\text{BH}}\sim\!10^9$M$_\odot$ based upon simulations (\citealt{Raouf2016}), although \cite{Buote2017} find no significant effect when including a SMBH of this mass in their model fits. In either case, the black hole in NGC 1600 is an order of magnitude larger in mass.

A direct comparison to one of the previously mentioned systems, MRK 1216, provides some useful insight. Both MRK 1216 and NGC 1600 have $t_{c}/t_{ff} < 20$ within their centers. While the region extends to roughly the scale radius r$_s = 12$ kpc for MRK 1216, the region of low $t_{c}/t_{ff}$ is within the central 3--5 kpc for NGC 1600, much lower then the calculated scale radius of r$_s = 15$ kpc. As pointed out in \cite{Buote2019}, for the region where $t_{c}/t_{ff}<20$, the precipitation-regulated feedback scenario says there should be multiphase gas present. This is true for NGC 1600 (\citealt{Runge2021}), but \cite{Buote2019} find no evidence for a cooler gas component in MRK 1216, suggesting that the condensation is being dampened due to a steeply rising entropy profile with a slope above a critical value of $\alpha=2/3$ (\citealt{Voit2017}). The entropy slope for NGC 1600 in this region is below the critical value, possibly explaining why we find a cooler gas component. It should also be noted that the temperature profile for MRK 1216 is peaked in the center while the temperature profile for NGC 1600 is cooler near the center with a mild increase within 1 kpc; suggesting a cool-core. This is not surprising considering that a study of 17 fossil systems by \cite{Bhar2016} shows that $\sim\!80\%$ are classified as cool-core; similar to the fraction for all galaxy clusters (\citealt{Hudson2010})

There is a caveat that should be noted though: being a fossil group does not guarantee a high concentration. Both \cite{Demo2010} and \cite{Pratt2016} have found normal concentration values for a number of fossil groups. However, the small sample size of fossil groups leaves out the possibility of making any assumptions about the population as a whole. Regardless of whether NGC 1600 is a fossil group or not, its high concentration value makes it an outlier among galaxy groups.

\subsection{AGN Feedback}

In order to investigate whether AGN feedback has any effect on the concentration parameter, we break the observation into four quadrants (See Fig.~\ref{fig:image_temp_compare}). As can be observed from the X-ray image, the East quadrant is relatively devoid of any evidence of recent AGN activity, while the other quadrants show varying degrees of disturbances. We repeat the same process outlined previously for each individual quadrant using only the \textit{Chandra} observations; increasing the radial bin-widths for sufficient counts per bin. After breaking up the data into four quadrants, we find the data quality is no longer sufficient to accurately be fit by a double-$\beta$ model; therefore, we switch to a single-$\beta$ model fit (only the first term in Eq. 4). The temperature profiles, assuming an NFW model, are shown on the right hand side of Figure~\ref{fig:image_temp_compare}. For the inner 1 kpc region, both the North and South quadrants show a slightly increasing temperature towards the center; while the East and West quadrants exhibit a decreasing temperature in the central region.
\begin{table}
    \centering
        \begin{tabular}{r @{\hspace{5pt}\vline\hspace{5pt}} c c @{\hspace{5pt}\vline\hspace{5pt}} c l}
        \toprule
        &\multicolumn{2}{c|}{$c_{200}$} & \multicolumn{2}{c|}{$r_{200}$}\\
        &\multicolumn{2}{c}{}&\multicolumn{2}{c}{(kpc)}\\
        Region & Best & $\Bar{c}_{200}$ & Best & \multicolumn{1}{c}{$\Bar{r}_{200}$}\\
        \toprule
    North & 27.5 & $25.5\pm3.9$ & 736.2 & $792\pm127$ \\
    East & 43.7 & $43.5\pm4.3$ & 452.9 & $458\pm33$ \\
    South & 39.8 & $41.3\pm4.7$ & 483.1 & $475\pm37$ \\
    West & 34.7 & $35.2\pm6.0$ & 552.1 & $551\pm60$ \\
    \toprule
        \end{tabular}
        \caption{Fitting results of $c_{200}$ and $r_{200}$ for the quadrants of NGC 1600 using \textit{Chandra} and a single-$\beta$ model. }
    \label{table:Quadrant_fit}
\end{table}

A breakdown of the calculated $c_{200}$ and r$_{200}$ values is given in Table~\ref{table:Quadrant_fit}. It is clear that the region with the least visible AGN activity, East, has the highest concentration parameter coupled with the lowest r$_{200}$. This is in contrast with the North region which has the lowest concentration value and the highest r$_{200}$ value, which is possibly due to AGN activity.

\begin{figure}
    \centering
    \includegraphics[width=0.49\textwidth]{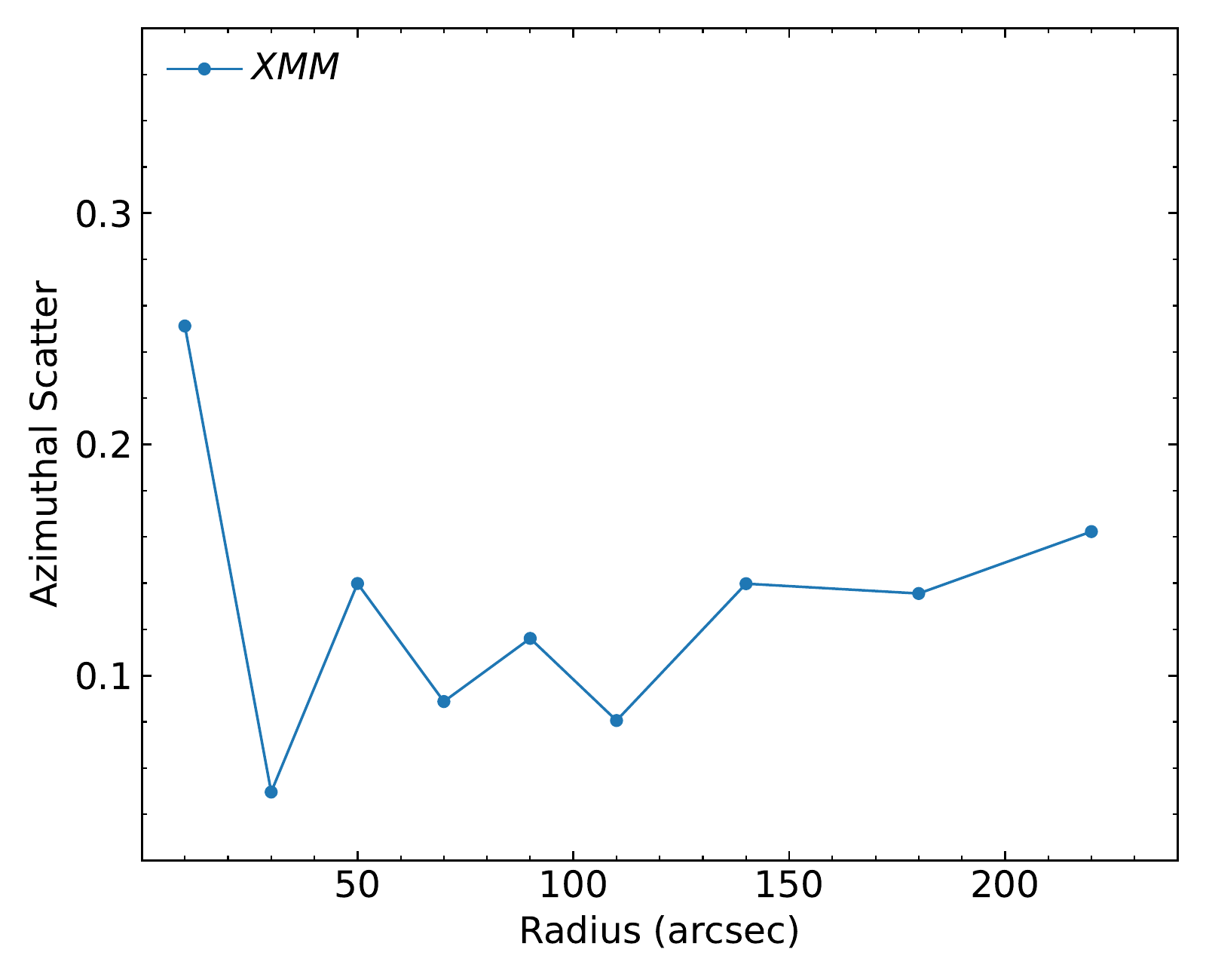}
    \caption{The azimuthal scatter for the surface brightness of NGC 1600 based on XMM data. A small value for the scatter indicates no significant deviation from spherical symmetry. The high central value is attributed to AGN feedback.}
    \label{fig:sb_scatter}
\end{figure}

To further quantify the asymmetry of the group, we calculated how the azimuthal scatter varies with distance from the core (\citealt{Vazza2011,Ghira2018,Mira2020}). The azimuthal scatter at radius $r$ is defined as  
\begin{equation}
\sigma_q(r) = \sqrt{\frac{1}{N}\sum_{i=1}^{N}\left ( \frac{q_i(r) - \overline{q}(r)}{ \overline{q}(r)} \right )^{2}}, 
\end{equation}
where N is the number of sectors, $q_i(r)$ is the measured quantity for sector $i$, and $\overline{q}(r)$ is the azimuthally averaged profile. To investigate the scatter at larger radii, we use the surface brightness measurements from XMM in the four quadrants previously mentioned. The azimuthal scatter is shown in Fig.~\ref{fig:sb_scatter}. The scatter peaks in the central region due to AGN feedback before decreasing and flattening as the radius increases. Therefore, the dominant source of asymmetry is the AGN, and its effect is limited to the central regions. NGC 1600 is a relaxed, cool-core system with a low central entropy ($\sim\!5$ keV cm$^2$, \citealt{Runge2021}), and has not undergone any merging recently, leaving it undisturbed. These results are consistent with studies of the pressure profiles of cool-core systems (e.g. \citealt{Arnaud2010}) which find that the scatter is highest in the center of cool-core clusters/groups due to AGN feedback, but becomes very low outside $\sim\!0.1r_{500}$, such that the pressure profile has low scatter and has a universal shape.

\section{Conclusions}

Through the use of deep \textit{Chandra} observations with an additional \textit{XMM--Newton} observations, we have measured a very high concentration for NGC 1600. By accounting for the mass of the black hole, the stellar mass, and the radially decreasing metal abundance, we estimate a concentration value as low as c$_{200}=26.7\pm1.4$ and c$_{200}=19.7\pm1.3$ for an NFW and Einasto DM profile respectively.

Such a high concentration is similar to the values that Buote finds for MRK 1216 and NGC 6482 ($c_{200}=30.4\pm4.3$ and $c_{200}=32.2\pm7.1$ respectively). In all cases, they are extreme outliers of the $\Lambda$CDM c$_{200}$-M$_{200}$ relation. While NGC 1600 does have a comparable c$_{200}$, its mass, M$_{200}\!=\!(2.0\pm0.2)\times10^{13}$ M$_\odot$, is four times larger than MRK 1216 or NGC 6482, both on the order of $5\times10^{12}$ M$_\odot$.

Despite not being considered a fossil group, arguments have been made to identify NGC 1600 as such. As outlined in \cite{Smith2008}, NGC 1600 does have many of the characteristics of a fossil group but lacks the high X-ray luminosity usually associated with one, which may be due to feedback from the $17\times10^{9}$ M$_\odot$ black hole. Our study reveals that NGC 1600 has similar DM halo properties of two well known fossil groups, MRK 1216 and NGC 6482. It also further confirms the relation that c$_{200}$ decreases as M$_{200}$ increases, a key prediction by CDM models. 

Regardless of being a fossil group or not, NGC 1600 is part of a small group of objects that have very high concentration values. More extensive research is needed to find and examine these outliers in order to determine what is the root cause of them to exhibit such high values.

\section*{Acknowledgements}
We thank the referee for helpful comments that improved the paper. JR, SAW and MSM acknowledge support from Chandra grant GO9-20073X. This work is based on observations obtained
with the Chandra observatory, a NASA mission.

\section*{Data Availability}
The Chandra Data Archive stores the data used in this paper. The \textit{Chandra} data were processed using the \textit{Chandra} Interactive Analysis of Observations (CIAO) software. The \textit{XMM--Newton} Science Archive (XSA) stores the archival data used in this paper, from which the data are publicly available for download. The \textit{XMM} data were processed using the \textit{XMM--Newton} Science Analysis System (\textsc{sas}). The software packages \textsc{heasoft} and \textsc{xspec} were used, and these can be downloaded from the High Energy Astrophysics Science Archive Research Centre (HEASARC) software web page.

\bibliography{reflib}

\begin{thebibliography}{}

\bibitem[\protect\citeauthoryear{{Arnaud}, {Pratt}, {Piffaretti},
  {B{\"o}hringer}, {Croston} \& {Pointecouteau}}{{Arnaud}
  et~al.}{2010}]{Arnaud2010}
{Arnaud} M.,  {Pratt} G.~W.,  {Piffaretti} R.,  {B{\"o}hringer} H.,  {Croston}
  J.~H.,    {Pointecouteau} E.,  2010, \aap, 517, A92

\bibitem[\protect\citeauthoryear{{Bharadwaj}, {Reiprich}, {Sanders} \&
  {Schellenberger}}{{Bharadwaj} et~al.}{2016}]{Bhar2016}
{Bharadwaj} V.,  {Reiprich} T.~H.,  {Sanders} J.~S.,    {Schellenberger} G.,
  2016, \aap, 585, A125

\bibitem[\protect\citeauthoryear{{Bhattacharya}, {Habib}, {Heitmann} \&
  {Vikhlinin}}{{Bhattacharya} et~al.}{2013}]{Bhattacharya2013}
{Bhattacharya} S.,  {Habib} S.,  {Heitmann} K.,    {Vikhlinin} A.,  2013, \apj,
  766, 32

\bibitem[\protect\citeauthoryear{{Bullock}, {Kolatt}, {Sigad}, {Somerville},
  {Kravtsov}, {Klypin}, {Primack} \& {Dekel}}{{Bullock}
  et~al.}{2001}]{Bullock2001}
{Bullock} J.~S.,  {Kolatt} T.~S.,  {Sigad} Y.,  {Somerville} R.~S.,  {Kravtsov}
  A.~V.,  {Klypin} A.~A.,  {Primack} J.~R.,    {Dekel} A.,  2001, \mnras, 321,
  559

\bibitem[\protect\citeauthoryear{{Buote}}{{Buote}}{2017}]{Buote2017}
{Buote} D.~A.,  2017, \apj, 834, 164

\bibitem[\protect\citeauthoryear{{Buote} \& {Barth}}{{Buote} \&
  {Barth}}{2019}]{Buote2019}
{Buote} D.~A.,  {Barth} A.~J.,  2019, \apj, 877, 91

\bibitem[\protect\citeauthoryear{{Buote}, {Gastaldello}, {Humphrey},
  {Zappacosta}, {Bullock}, {Brighenti} \& {Mathews}}{{Buote}
  et~al.}{2007}]{Buote2007}
{Buote} D.~A.,  {Gastaldello} F.,  {Humphrey} P.~J.,  {Zappacosta} L.,
  {Bullock} J.~S.,  {Brighenti} F.,    {Mathews} W.~G.,  2007, \apj, 664, 123

\bibitem[\protect\citeauthoryear{{D{\'e}mocl{\`e}s}, {Pratt}, {Pierini},
  {Arnaud}, {Zibetti} \& {D'Onghia}}{{D{\'e}mocl{\`e}s}
  et~al.}{2010}]{Demo2010}
{D{\'e}mocl{\`e}s} J.,  {Pratt} G.~W.,  {Pierini} D.,  {Arnaud} M.,  {Zibetti}
  S.,    {D'Onghia} E.,  2010, \aap, 517, A52

\bibitem[\protect\citeauthoryear{{Dickey} \& {Lockman}}{{Dickey} \&
  {Lockman}}{1990}]{DL90}
{Dickey} J.~M.,  {Lockman} F.~J.,  1990, \araa, 28, 215

\bibitem[\protect\citeauthoryear{{Dutton} \& {Macci{\`o}}}{{Dutton} \&
  {Macci{\`o}}}{2014}]{Dutton2014}
{Dutton} A.~A.,  {Macci{\`o}} A.~V.,  2014, \mnras, 441, 3359

\bibitem[\protect\citeauthoryear{{Einasto}}{{Einasto}}{1965}]{Einasto65}
{Einasto} J.,  1965, Trudy Astrofizicheskogo Instituta Alma-Ata, 5, 87

\bibitem[\protect\citeauthoryear{{Ettori}, {Gastaldello}, {Leccardi},
  {Molendi}, {Rossetti}, {Buote} \& {Meneghetti}}{{Ettori}
  et~al.}{2010}]{Ettori2010}
{Ettori} S.,  {Gastaldello} F.,  {Leccardi} A.,  {Molendi} S.,  {Rossetti} M.,
  {Buote} D.,    {Meneghetti} M.,  2010, \aap, 524, A68

\bibitem[\protect\citeauthoryear{{Ghirardini}, {Ettori}, {Eckert}, {Molendi},
  {Gastaldello}, {Pointecouteau}, {Hurier} \& {Bourdin}}{{Ghirardini}
  et~al.}{2018}]{Ghira2018}
{Ghirardini} V.,  {Ettori} S.,  {Eckert} D.,  {Molendi} S.,  {Gastaldello} F.,
  {Pointecouteau} E.,  {Hurier} G.,    {Bourdin} H.,  2018, \aap, 614, A7

\bibitem[\protect\citeauthoryear{{Harrison}, {Miller}, {Richards},
  {Lloyd-Davies}, {Hoyle}, {Romer}, {Mehrtens}, {Hilton}, {Stott}, {Capozzi},
  {Collins}, {Deadman}, {Liddle}, {Sahl{\'e}n}, {Stanford} \&
  {Viana}}{{Harrison} et~al.}{2012}]{Harrison2012}
{Harrison} C.~D.,  {Miller} C.~J.,  {Richards} J.~W.,  {Lloyd-Davies} E.~J.,
  {Hoyle} B.,  {Romer} A.~K.,  {Mehrtens} N.,  {Hilton} M.,  {Stott} J.~P.,
  {Capozzi} D.,  {Collins} C.~A.,  {Deadman} P.-J.,  {Liddle} A.~R.,
  {Sahl{\'e}n} M.,  {Stanford} S.~A.,    {Viana} P. T.~P.,  2012, \apj, 752, 12

\bibitem[\protect\citeauthoryear{{Hudson}, {Mittal}, {Reiprich}, {Nulsen},
  {Andernach} \& {Sarazin}}{{Hudson} et~al.}{2010}]{Hudson2010}
{Hudson} D.~S.,  {Mittal} R.,  {Reiprich} T.~H.,  {Nulsen} P.~E.~J.,
  {Andernach} H.,    {Sarazin} C.~L.,  2010, \aap, 513, A37

\bibitem[\protect\citeauthoryear{{Humphrey} \& {Buote}}{{Humphrey} \&
  {Buote}}{2006}]{Humphrey2006}
{Humphrey} P.~J.,  {Buote} D.~A.,  2006, \apj, 639, 136

\bibitem[\protect\citeauthoryear{{Humphrey}, {Buote}, {Brighenti}, {Flohic},
  {Gastaldello} \& {Mathews}}{{Humphrey} et~al.}{2012}]{Humphrey2012}
{Humphrey} P.~J.,  {Buote} D.~A.,  {Brighenti} F.,  {Flohic} H. M.~L.~G.,
  {Gastaldello} F.,    {Mathews} W.~G.,  2012, \apj, 748, 11

\bibitem[\protect\citeauthoryear{{Humphrey}, {Buote}, {Canizares}, {Fabian} \&
  {Miller}}{{Humphrey} et~al.}{2011}]{Humphrey2011}
{Humphrey} P.~J.,  {Buote} D.~A.,  {Canizares} C.~R.,  {Fabian} A.~C.,
  {Miller} J.~M.,  2011, \apj, 729, 53

\bibitem[\protect\citeauthoryear{{Jarrett}, {Chester}, {Cutri}, {Schneider},
  {Skrutskie} \& {Huchra}}{{Jarrett} et~al.}{2000}]{Jarrett2000}
{Jarrett} T.~H.,  {Chester} T.,  {Cutri} R.,  {Schneider} S.,  {Skrutskie} M.,
    {Huchra} J.~P.,  2000, \aj, 119, 2498

\bibitem[\protect\citeauthoryear{{Jones}, {Ponman}, {Horton}, {Babul},
  {Ebeling} \& {Burke}}{{Jones} et~al.}{2003}]{Jones2003}
{Jones} L.~R.,  {Ponman} T.~J.,  {Horton} A.,  {Babul} A.,  {Ebeling} H.,
  {Burke} D.~J.,  2003, \mnras, 343, 627

\bibitem[\protect\citeauthoryear{{Ma}, {Greene}, {McConnell}, {Janish},
  {Blakeslee}, {Thomas} \& {Murphy}}{{Ma} et~al.}{2014}]{Ma2014}
{Ma} C.-P.,  {Greene} J.~E.,  {McConnell} N.,  {Janish} R.,  {Blakeslee} J.~P.,
   {Thomas} J.,    {Murphy} J.~D.,  2014, \apj, 795, 158

\bibitem[\protect\citeauthoryear{{Macci{\`o}}, {Dutton}, {van den Bosch},
  {Moore}, {Potter} \& {Stadel}}{{Macci{\`o}} et~al.}{2007}]{Macci2007}
{Macci{\`o}} A.~V.,  {Dutton} A.~A.,  {van den Bosch} F.~C.,  {Moore} B.,
  {Potter} D.,    {Stadel} J.,  2007, \mnras, 378, 55

\bibitem[\protect\citeauthoryear{{Mernier}, {de Plaa}, {Kaastra}, {Zhang},
  {Akamatsu}, {Gu}, {Kosec}, {Mao}, {Pinto}, {Reiprich}, {Sanders},
  {Simionescu} \& {Werner}}{{Mernier} et~al.}{2017}]{Mernier2017}
{Mernier} F.,  {de Plaa} J.,  {Kaastra} J.~S.,  {Zhang} Y.~Y.,  {Akamatsu} H.,
  {Gu} L.,  {Kosec} P.,  {Mao} J.,  {Pinto} C.,  {Reiprich} T.~H.,  {Sanders}
  J.~S.,  {Simionescu} A.,    {Werner} N.,  2017, \aap, 603, A80

\bibitem[\protect\citeauthoryear{{Mirakhor} \& {Walker}}{{Mirakhor} \&
  {Walker}}{2020}]{Mira2020}
{Mirakhor} M.~S.,  {Walker} S.~A.,  2020, \mnras, 497, 3204

\bibitem[\protect\citeauthoryear{{Navarro}, {Frenk} \& {White}}{{Navarro}
  et~al.}{1996}]{Navarro96}
{Navarro} J.~F.,  {Frenk} C.~S.,    {White} S. D.~M.,  1996, \apj, 462, 563

\bibitem[\protect\citeauthoryear{{Navarro}, {Frenk} \& {White}}{{Navarro}
  et~al.}{1997}]{NFW97}
{Navarro} J.~F.,  {Frenk} C.~S.,    {White} S. D.~M.,  1997, \apj, 490, 493

\bibitem[\protect\citeauthoryear{{Ponman}, {Allan}, {Jones}, {Merrifield},
  {McHardy}, {Lehto} \& {Luppino}}{{Ponman} et~al.}{1994}]{Ponman1994}
{Ponman} T.~J.,  {Allan} D.~J.,  {Jones} L.~R.,  {Merrifield} M.,  {McHardy}
  I.~M.,  {Lehto} H.~J.,    {Luppino} G.~A.,  1994, \nat, 369, 462

\bibitem[\protect\citeauthoryear{{Pratt}, {Pointecouteau}, {Arnaud} \& {van der
  Burg}}{{Pratt} et~al.}{2016}]{Pratt2016}
{Pratt} G.~W.,  {Pointecouteau} E.,  {Arnaud} M.,    {van der Burg} R.~F.~J.,
  2016, \aap, 590, L1

\bibitem[\protect\citeauthoryear{{Raouf}, {Khosroshahi} \& {Dariush}}{{Raouf}
  et~al.}{2016}]{Raouf2016}
{Raouf} M.,  {Khosroshahi} H.~G.,    {Dariush} A.,  2016, \apj, 824, 140

\bibitem[\protect\citeauthoryear{{Rasmussen} \& {Ponman}}{{Rasmussen} \&
  {Ponman}}{2009}]{Rasmussen2009}
{Rasmussen} J.,  {Ponman} T.~J.,  2009, \mnras, 399, 239

\bibitem[\protect\citeauthoryear{{Runge} \& {Walker}}{{Runge} \&
  {Walker}}{2021}]{Runge2021}
{Runge} J.,  {Walker} S.~A.,  2021, \mnras

\bibitem[\protect\citeauthoryear{{Sanders}, {Fabian}, {Russell} \&
  {Walker}}{{Sanders} et~al.}{2018}]{Sanders2018}
{Sanders} J.~S.,  {Fabian} A.~C.,  {Russell} H.~R.,    {Walker} S.~A.,  2018,
  \mnras, 474, 1065

\bibitem[\protect\citeauthoryear{{Schmidt} \& {Allen}}{{Schmidt} \&
  {Allen}}{2007}]{Schmidt2007}
{Schmidt} R.~W.,  {Allen} S.~W.,  2007, \mnras, 379, 209

\bibitem[\protect\citeauthoryear{{Sivakoff}, {Sarazin} \& {Carlin}}{{Sivakoff}
  et~al.}{2004}]{Sivakoff2004}
{Sivakoff} G.~R.,  {Sarazin} C.~L.,    {Carlin} J.~L.,  2004, \apj, 617, 262

\bibitem[\protect\citeauthoryear{{Skrutskie}, {Cutri}, {Stiening}, {Weinberg},
  {Schneider}, {Carpenter}, {Beichman}, {Capps}, {Chester}, {Elias}, {Huchra},
  {Liebert}, {Lonsdale}, {Monet}, {Price}, {Seitzer}, {Jarrett}, {Kirkpatrick},
  {Gizis}, {Howard}, {Evans}, {Fowler}, {Fullmer}, {Hurt}, {Light}, {Kopan},
  {Marsh}, {McCallon}, {Tam}, {Van Dyk} \& {Wheelock}}{{Skrutskie}
  et~al.}{2006}]{Skrutskie2006}
{Skrutskie} M.~F.,  {Cutri} R.~M.,  {Stiening} R.,  {Weinberg} M.~D.,
  {Schneider} S.,  {Carpenter} J.~M.,  {Beichman} C.,  {Capps} R.,  {Chester}
  T.,  {Elias} J.,  {Huchra} J.,  {Liebert} J.,  {Lonsdale} C.,  {Monet} D.~G.,
   {Price} S.,  {Seitzer} P.,  {Jarrett} T.,  {Kirkpatrick} J.~D.,  {Gizis}
  J.~E.,  {Howard} E.,  {Evans} T.,  {Fowler} J.,  {Fullmer} L.,  {Hurt} R.,
  {Light} R.,  {Kopan} E.~L.,  {Marsh} K.~A.,  {McCallon} H.~L.,  {Tam} R.,
  {Van Dyk} S.,    {Wheelock} S.,  2006, \aj, 131, 1163

\bibitem[\protect\citeauthoryear{{Smith}, {Mart{\'\i}nez},
  {Fern{\'a}ndez-Soto}, {Ballesteros} \& {Ortiz-Gil}}{{Smith}
  et~al.}{2008}]{Smith2008}
{Smith} R.~M.,  {Mart{\'\i}nez} V.~J.,  {Fern{\'a}ndez-Soto} A.,  {Ballesteros}
  F.~J.,    {Ortiz-Gil} A.,  2008, \apj, 679, 420

\bibitem[\protect\citeauthoryear{{Thomas}, {Ma}, {McConnell}, {Greene},
  {Blakeslee} \& {Janish}}{{Thomas} et~al.}{2016}]{Thomas2016}
{Thomas} J.,  {Ma} C.-P.,  {McConnell} N.~J.,  {Greene} J.~E.,  {Blakeslee}
  J.~P.,    {Janish} R.,  2016, \nat, 532, 340

\bibitem[\protect\citeauthoryear{{Vazza}, {Roncarelli}, {Ettori} \&
  {Dolag}}{{Vazza} et~al.}{2011}]{Vazza2011}
{Vazza} F.,  {Roncarelli} M.,  {Ettori} S.,    {Dolag} K.,  2011, \mnras, 413,
  2305

\bibitem[\protect\citeauthoryear{{Vikhlinin}, {Kravtsov}, {Forman}, {Jones},
  {Markevitch}, {Murray} \& {Van Speybroeck}}{{Vikhlinin}
  et~al.}{2006}]{Vik2006}
{Vikhlinin} A.,  {Kravtsov} A.,  {Forman} W.,  {Jones} C.,  {Markevitch} M.,
  {Murray} S.~S.,    {Van Speybroeck} L.,  2006, \apj, 640, 691

\bibitem[\protect\citeauthoryear{{Voit}, {Meece}, {Li}, {O'Shea}, {Bryan} \&
  {Donahue}}{{Voit} et~al.}{2017}]{Voit2017}
{Voit} G.~M.,  {Meece} G.,  {Li} Y.,  {O'Shea} B.~W.,  {Bryan} G.~L.,
  {Donahue} M.,  2017, \apj, 845, 80

\bibitem[\protect\citeauthoryear{{Walsh}, {van den Bosch}, {Gebhardt},
  {Y{\i}ld{\i}r{\i}m}, {G{\"u}ltekin}, {Husemann} \& {Richstone}}{{Walsh}
  et~al.}{2017}]{Walsh2017}
{Walsh} J.~L.,  {van den Bosch} R. C.~E.,  {Gebhardt} K.,  {Y{\i}ld{\i}r{\i}m}
  A.,  {G{\"u}ltekin} K.,  {Husemann} B.,    {Richstone} D.~O.,  2017, \apj,
  835, 208

\end{thebibliography}
\end{document}